\documentstyle[12pt,epsfig]{article} 

\begin{document}  
\vspace*{-2cm}  
\renewcommand{\thefootnote}{\fnsymbol{footnote}}  
\vskip 45pt  
\begin{center}  
{\Large \bf Electroweak Sudakov corrections in the MSSM} \\
\vspace{1.2cm} 

\bf{M. Beccaria\footnote{Dipartimento di Fisica, Universit\`a di
Lecce,INFN, Sezione di Lecce, Italy}, M. Melles\footnote{Paul Scherrer Institute (PSI), Switzerland}
F.M. Renard\footnote{Physique
Math\'{e}matique et Th\'{e}orique, Universit\'{e} Montpellier II, France} and C. Verzegnassi\footnote{
Dipartimento di Fisica Teorica, Universit\`a di Trieste, Italy}\\
\vspace{0.4cm}
}

\vspace{20pt}  
\begin{abstract}
For superpartner masses not much heavier than the weak scale $M=M_{\rm W}$, large logarithmic
corrections of the Sudakov type arise at TeV energies. In this paper we summarize recent results
of supersymmetric (susy) electroweak radiative corrections for sfermion and 
charged Higgs production at $e^+e^-$
colliders in the MSSM. The results are given to subleading logarithmic (SL) accuracy to all
orders in perturbation theory in the ``light'' susy-mass scenario. Prospects for the
determination of $\tan \beta \geq 10$ are discussed which is independent of soft susy-breaking
terms to SL accuracy.
\end{abstract}
\end{center}  
\vskip12pt

\setcounter{footnote}{0}  
\renewcommand{\thefootnote}{\arabic{footnote}}  
  
Recently, there has been considerable interest in the high energy limit of the electroweak Standard Model
(SM) \cite{flmm,brv1,m1,m2,m3,m4,kps,kmps,dp1,dp2,ccc1,bw,hkk,m6}. 
Those studies have concluded that at TeV energies virtual corrections of the Sudakov type
are very large and higher order resummations are necessary to reach the desired percentile accuracy
of future TeV linear colliders. Presently, a full subleading logarithmic (SL) approach exists in terms of the
infrared evolution equation method \cite{kl} for arbitrary processes to all orders \cite{flmm,m5,habil}. 
Further studies for massless
fermions indicate that also sub-subleading angular terms can be large and need to be included
at least through the two loop level \cite{kmps}.

In general, new physics responsible for electroweak symmetry breaking is expected in the TeV regime and
the minimal supersymmetric SM (MSSM) remains an attractive candidate. If supersymmetry is relevant
to the so called hierarchy problem, then the masses of the new superpartners cannot be much heavier
than the weak scale $M \equiv M_{\rm W} \sim M_{\rm Z}$. In such a ``light'' susy mass scenario,
similarly large radiative corrections can be expected as in the SM at TeV energies. At one loop
this was confirmed by several works of the last few years \cite{brv2,brv3}. 

As in the SM, large corrections at the one loop level indicate that higher order contributions need
to be included, both for the consistency of perturbation theory as well as the precision goals at
a future TeV linear collider. In Ref. \cite{bmrv} we have presented results of higher order 
electroweak corrections
to scalar production in the context of the MSSM. 

The leading double (DL) and subleading angular dependent corrections originate only from the exchange
of spin 1 gauge bosons and are therefore in principle identical to those of the SM. This is
due to the assumption of softly broken supersymmetry and the identity of the gauge couplings
in those theories between the gauge bosons and the SM particles with those of the 
gauge bosons and the accompanying sparticles.
Novel contributions compared to the SM arise on the SL level, however,  
both from the gauge as well as the Yukawa sector through
the novel particle content. 

As in the SM, the SL corrections were found to
exponentiate in operator form on the n-particle space with rotated Born-matrix elements \cite{dp1,m5}.
This conclusion, obtained in the effective theories beforehand (in Refs. \cite{kps} and \cite{kmps} at first for
massless fermion production processes and in Ref. \cite{m5} for arbitrary electroweak processes) has
recently been confirmed by explicit calculation in terms of the physical fields for the angle
dependent SL corrections \cite{dmp}.
The universal, process independent SL corrections exponentiate due to Ward identities in both
the gauge as well as the Yukawa sector \cite{m3,bmrv}.

A slight complication is given by the so-called SL-RG dependent terms \cite{m4,ms}, where anomalous dimension
terms proportional to the MSSM $\beta$-functions arise. In the following we give these corrections
for the case that the full particle content contributes which, however, in a real TeV collider 
environment might not be the case. Since the scale of these terms is $m_s$, the mass-scale of
the superpartners, and since we assume $m_s \sim M$, the sub-subleading corrections 
of the type ${\cal O} \left( \alpha^2 \log \frac{m_s^2}{M^2} \log^2 \frac{s}{M^2} \right)$, which originate
from this problem, are negligible. Similar sub-subleading terms could also modify the SL gauge contributions
if some susy particles are heavier.
Nevertheless for clarity we distinguish these two scales below and also
give the results only for the case of a heavy photon with $\lambda=M$. The omitted QED type corrections
must be included via matching as in the SM \cite{flmm,habil}.

Under the above assumptions we find for the production of sfermions (${\tilde f}$) in $e^+e^-$ collisions the
following all orders MSSM corrections to SL accuracy relative to the Born cross section:
\begin{eqnarray}
d \sigma^{\rm SL}_{e^+_{\alpha} e^-_{\alpha} \longrightarrow {\overline {\tilde f}}_{\beta}
{\tilde f}_{\beta}}
&=& d \sigma^{\rm Born}_{e^+_{\alpha} e^-_{\alpha} \longrightarrow {\overline {\tilde f}}_{\beta}
{\tilde f}_{\beta}} \times \nonumber \\ && \exp \left\{ - \frac{g^2(m_s^2)
I_{e^-_{\alpha}}(I_{e^-_{\alpha}}+1)}{8 \pi^2 } \left[ \log^2 \frac{q^2}{M^2}
- \frac{1}{3} {\tilde \beta_0} \frac{g^2(m_s^2) }{4 \pi^2 } \log^3 \frac{q^2}{m_s^2} \right] \right. \nonumber \\
&& -\frac{{g^\prime}^2(m_s^2) Y^2_{e^-_{\alpha}}}{32 \pi^2 } \left[
\log^2 \frac{q^2}{M^2}- \frac{1}{3} {\tilde \beta_0}^\prime \frac{{g^\prime}^2(m_s^2)
}{4 \pi^2 } \log^3 \frac{q^2}{m_s^2} \right]
\nonumber \\
&& + \left( \frac{ g^2(m_s^2)}{8 \pi^2} I_{e^-_{\alpha}}(I_{e^-_{\alpha}}+1)+
\frac{ {g^\prime}^2(m_s^2)}{8 \pi^2}\frac{Y^2_{e^-_{\alpha}}}{4} \right)  2 \log \frac{q^2}{M^2}
\nonumber \\ &&
- \frac{g^2(m_s^2)
I_{{\tilde f}_{\beta}}(I_{{\tilde f}_{\beta}}+1)}{8 \pi^2 } \left[ \log^2 \frac{q^2}{M^2}
- \frac{1}{3} {\tilde \beta_0} \frac{g^2(m_s^2) }{4 \pi^2 } \log^3 \frac{q^2}{m_s^2} \right]  \nonumber \\
&& -\frac{{g^\prime}^2(m_s^2) Y^2_{{\tilde f}_{\beta}}}{32 \pi^2 } \left[
\log^2 \frac{q^2}{M^2}- \frac{1}{3} {\tilde \beta_0}^\prime \frac{{g^\prime}^2(m_s^2)
}{4 \pi^2 } \log^3 \frac{q^2}{m_s^2} \right]
\nonumber \\
&& + \left( \frac{ g^2(m_s^2)}{8 \pi^2} I_{{\tilde f}_{\beta}}(I_{{\tilde f}_{\beta}}+1)+
\frac{ {g^\prime}^2(m_s^2)}{8 \pi^2}\frac{Y^2_{{\tilde f}_{\beta}}}{4} \right)  2 \log \frac{q^2}{M^2}
\nonumber \\ &&
- \frac{ g^2(m_s^2)}{8 \pi^2} \left( \frac{1+\delta_{\beta,{\rm R}}}{2} \frac{{\hat m}
^2_{\tilde f}}{M^2} + \delta_{\beta,{\rm L}}
\frac{{\hat m}^2_{{\tilde f}^\prime}}{2 M^2} \right)
\log \frac{q^2}{m_s^2} \nonumber \\ &&
-\frac{g^2(m_s^2)}{8\pi^2} \log \frac{q^2}{M^2} \left[ \left( \tan^2 \theta_{\rm w} Y_{e^-_{\alpha}} Y_{
{\tilde f}_\beta}
+ 4 I^3_{e^-_{\alpha}} I^3_{{\tilde f}_\beta} \right) \log \frac{t}{u} \right. \nonumber \\ &&
\left. \left. + \frac{\delta_{\alpha, L} \delta_{\beta,L}}{\tan^2 \theta_{\rm w} Y_{e^-_{\alpha}} Y_{
{\tilde f}_\beta} /4
+ I^3_{e^-_{\alpha}} I^3_{{\tilde f}_\beta}} \left( \delta_{d,{\tilde f}} \log \frac{-t}{q^2} - 
\delta_{u,{\tilde f}}
\log \frac{-u}{q^2} \right)  \right] \right\}
\label{eq:sfSL}
\end{eqnarray}
where $I_j$ denotes the total weak isospin of the particle $j$, $Y_j$ its weak hypercharge
and at high $q^2$ the invariants are given by
$t=-\frac{q^2}{2} \left( 1-\cos \theta \right)$ and $u=-\frac{q^2}{2} \left( 1+\cos \theta \right)$.
The helicities are those of the fermions ($f$) whose superpartner is produced.
In addition we denote ${\hat m}_{\tilde f}=m_t / \sin \beta$
if ${\tilde f}= {\tilde t}$ and ${\hat m}_{\tilde f}=m_b / \cos \beta$ if ${\tilde f}={\tilde b}$. 
${\tilde f}^\prime$ denotes the corresponding
isopartner of ${\tilde f}$. For particles other than those belonging to the third family
of quarks/squarks, the Yukawa terms are negligible.
Eq. (\ref{eq:sfSL}) depends on the important parameter $\tan \beta = \frac{v_u}{v_d}$, the
ratio of the two vacuum expectation values, and displays
an exact supersymmetry in the sense that the same corrections are obtained for the fermionic
sector in the regime above the electroweak scale $M$.

Here we assume that the asymptotic MSSM $\beta$-functions can be used with
\begin{eqnarray}
{\tilde \beta}_0&=& \frac{3}{4} C_A- \frac{n_g}{2}-\frac{n_h}{8} \;\;,\;
{\tilde \beta}_0^\prime=-\frac{5}{6}n_g-\frac{n_h}{8} \label{eq:bMSSM} \\
g^2(q^2) &=& \frac{g^2(m_s^2)}{1+{\tilde \beta}_0 \frac{g^2
(m_s^2)}{4\pi^2}
\ln \frac{q^2}{m_s^2}} \;\;,\;
{g^\prime}^2 (q^2) = \frac{{g^\prime}^2 (m_s^2)}{1+{\tilde \beta}^\prime_0
\frac{{g^\prime}^2 (m_s^2)}{4\pi^2}
\ln \frac{q^2}{m_s^2}} \label{eq:arunMSSM}
\end{eqnarray}
where $C_A=2$, $n_g=3$ and $n_h=2$. In practice, one has to use the relevant numbers of active
particles in the loops. These terms correspond to the RG-SL corrections just as in the case
of the SM as discussed in Ref. \cite{m4}
but now with the MSSM particle spectrum contributing.
They originate only from RG terms within loops which without the RG contribution
would give a DL correction.
It should be noted that the one-loop RG corrections do not exponentiate and are omitted
in the above expressions. They are, however, completely determined by the renormalization group
in softly broken supersymmetric theories such as the MSSM
and sub-subleading at higher than one loop order.

In the case of charged Higgs production we have analogously:
\begin{eqnarray}
&& \!\!\!\!\!\!\!\!\!\!\!\!\!\!\!\!\!\!\!\!
d \sigma^{\rm SL}_{e^+_{\alpha} e^-_{\alpha} \longrightarrow H^+ H^-}
\!=\!\! d \sigma^{\rm Born}_{e^+_{\alpha} e^-_{\alpha} \longrightarrow H^+ H^-}
\times \nonumber \\ && \!\!\!\!\!\!\!\!\!\!\!\!\!\!\!\!\!\!\!\!
\exp \left\{ - \frac{g^2(m_s^2)}{8\pi^2} I_{e^-_{\alpha}} \left( I_{e^-_{\alpha}}+1
\right) \left[ 
\log^2 \frac{q^2}{M^2}
 - \frac{1}{3} {\tilde \beta_0} \frac{g^2(m_s^2) }{4 \pi^2 } \log^3 \frac{q^2}{m_s^2} 
\right] \right. \nonumber \\
&&\!\!\!\!\!\!\!\!\!\!\!\!\!\!\!\!\!\!\!\! -\frac{{g^\prime}^2(m_s^2) Y^2_{e^-_{\alpha}}}{32 \pi^2 }
\left[
\log^2 \frac{q^2}{M^2}- \frac{1}{3} {\tilde \beta_0}^\prime \frac{{g^\prime}^2(m_s^2)
}{4 \pi^2 } \log^3 \frac{q^2}{m_s^2}
\right] \nonumber \\
&&\!\!\!\!\!\!\!\!\!\!\!\!\!\!\!\!\!\!\!\! + \left( \frac{ g^2(m_s^2)}{8 \pi^2}
I_{e^-_{\alpha}} \left( I_{e^-_{\alpha}}+1 \right)+
\frac{ {g^\prime}^2(m_s^2)}{8 \pi^2} \frac{Y^2_{e^-_{\alpha}}}{4} \right)  2 \log \frac{q^2}{M^2}
\nonumber \\ &&\!\!\!\!\!\!\!\!\!\!\!\!\!\!\!\!\!\!\!\!
- \frac{g^2(m_s^2)}{8\pi^2} I_H \left( I_H+1
\right) \left[ 
\log^2 \frac{q^2}{M^2}
 - \frac{1}{3} {\tilde \beta_0} \frac{g^2(m_s^2) }{4 \pi^2 } \log^3 \frac{q^2}{m_s^2} 
\right] \nonumber \\
\!\!\!\!&&\!\!\!\!\!\!\!\!\!\!\!\!\!\!\!\!\!\!\!\! -\frac{{g^\prime}^2(m_s^2)
Y^2_H}{32 \pi^2 }
\left[
\log^2 \frac{q^2}{M^2}- \frac{1}{3} {\tilde \beta_0}^\prime \frac{{g^\prime}^2(m_s^2)
}{4 \pi^2 } \log^3 \frac{q^2}{m_s^2}
\right] \nonumber \\
\!\!\!\!&&\!\!\!\!\!\!\!\!\!\!\!\!\!\!\!\!\!\!\!\! + \left( \frac{ g^2(m_s^2)}{8 \pi^2}
I_H \left( I_H+1 \right)+
\frac{ {g^\prime}^2(m_s^2)}{8 \pi^2} \frac{Y^2_H}{4} \right)  2 \log \frac{q^2}{
M^2}
\nonumber \\ && \!\!\!\!\!\!\!\!\!\!\!\!\!\!\!\!\!\!\!\! 
- 3 \; \frac{g^2(m_s^2)}{32 \pi^2} \left[ \frac{m_t^2}{M^2} \cot^2 \beta + \frac{m_b^2}{M^2} \tan^2 \beta \right]
\log \frac{q^2}{m_s^2}
\nonumber \\ && \!\!\!\!\!\!\!\!\!\!\!\!\!\!\!\!\!\!\!\! \left.
-\frac{g^2 (m_s^2)}{4\pi^2} \log \frac{q^2}{M^2} \left[
\delta_{\alpha,{\rm L}} \left( \frac{1}{2 c^2_{\rm w}} \log \frac{t}{u} + 2 c^2_{\rm w}
\log \frac{-t}{q^2} \right)
+ \delta_{\alpha, {\rm R}} \tan^2 \theta_{\rm w}
\log \frac{t}{u}  \right] \right\} \label{eq:Hang}
\end{eqnarray}
It should be noted here that the Yukawa terms proportional to $\tan \beta$ are quite large
due to the additional factor of $3=N_C$ from the quark loops \cite{bmrv}. Overall, both Yukawa contributions
in Eq. (\ref{eq:sfSL}) and (\ref{eq:Hang}) reinforce the Sudakov suppression factor of the leading
double logarithmic terms. In addition, the universal positive SL gauge terms are identical (and compared
to the SM smaller) for spin $\frac{1}{2}$
and spin 0 particles due to the exact supersymmetry present at high energies. All supersymmetry
breaking terms are constants and thus beyond our level of approximation.

\begin{figure}
\centering
\epsfig{file=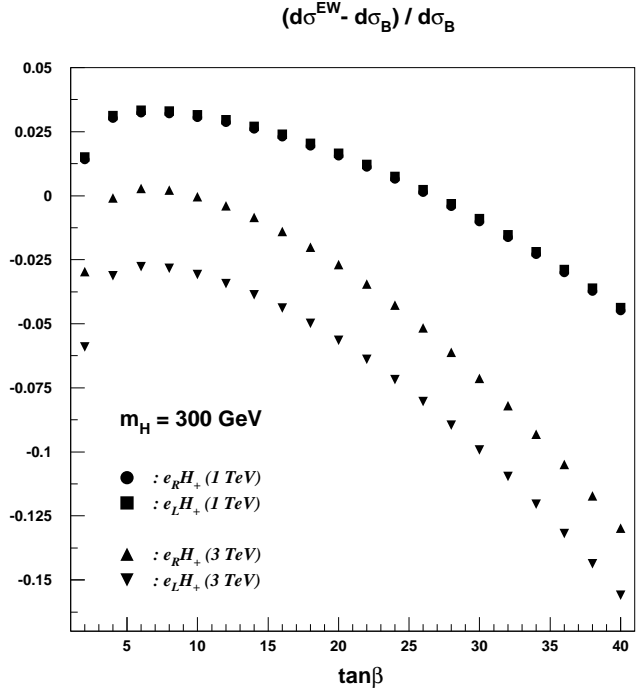,width=10cm}
\caption{The dependence of the relative corrections to the Born cross section for charged Higgs
production ($m_H=300$ GeV) as a function of $\tan \beta$ at $90^o$ scattering angle. 
It can be seen that for large values of $\tan \beta$, the Sudakov suppression is enhanced.}
\label{fig:Htanb}
\end{figure}

In Fig. \ref{fig:Htanb} we display the $\tan \beta$ dependence of the relative corrections to
charged Higgs production for $m_H=300$ GeV. It demonstrates that the relative $\tan \beta$ dependence of
the resummed corrections alone accounts for over 10 \% at 3 TeV  and over 5 \% at 1 TeV
in the window $10 \leq \tan \beta \leq 40$.

Since the precise measurement of large values of $\tan \beta$ at the LHC is not straightforward \cite{djou}
or strongly model dependent \cite{dre,maj,hin}, it is interesting to investigate if 
this significant dependence on
$\tan \beta$ through virtual corrections can be used for an independent measurement. 
In this context we envision a series of $N$ precise measurements 
at $ \sqrt{q_1^2}, \sqrt{q_2^2}, \dots,\sqrt{q_N^2}$ of various cross sections and introduce
the one loop quantity $\epsilon(q^2)$ as
the difference between measurements and the theoretical asymptotic DL and SL logarithms
of {\it gauge} origin. Then its asymptotic expansion is 
given by 
\begin{equation}
\epsilon(q^2) \equiv \frac{\alpha}{4\pi}\ F(\tan\beta) \ln\frac{q^2}{m_s^2} + G +
{\cal O}\left(\frac{M^2}{q^2}\right)
\end{equation}
Here it is crucial to note that $G$ is a constant which depends on mass ratios and that the function
$F(\tan \beta)$ does {\it not} depend on soft breaking terms.
Then we can eliminate the influence of the mass terms by subtraction via
\begin{equation}
\delta_i \equiv \epsilon(q_i^2)-\epsilon(q_1^2) = F(\tan\beta^*) \ \ln\frac{q_i^2}{q_1^2}
\end{equation}
where $\tan\beta^*$ is the {\em true} unknown value that describes the
experimental measurements.

\begin{figure}
\centering
\epsfig{file=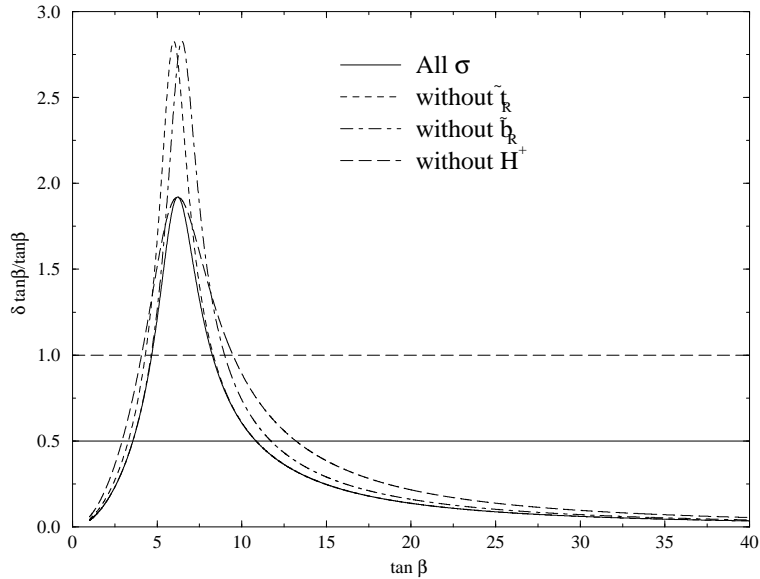,width=10cm,angle=-90}
\caption{The relative error in $\tan \beta$ after performing 10 measurements with a relative
accuracy of 1 \%. The curves can be improved if a higher precision could be reached.}
\label{fig:tanb}
\end{figure}

Performing a minimizing $\chi^2$ analysis leads to the results for the accuracy of the
$\tan \beta$ determination depicted in Fig. \ref{fig:tanb} as a function of 
the size of $\tan \beta$ and assuming a 1 \% accuracy of 10 measurements of the various
scalar on-shell production cross sections 
in the range between 0.8 and 3 GeV \cite{bmrv}.

It is clearly visible that a 50 \% measurement is possible for $\tan \beta \geq 10$,
a 25 \% determination for $\tan \beta \geq 15$ and a few percent measurement for
$\tan \beta \geq 25$.
Thus, the virtual radiative MSSM corrections are not only crucial for precision measurements
at a TeV linear collider at one and two loop order, they also contain valuable information
about the important Higgs sector parameter $\tan \beta$.

In particular we emphasize again that this gauge invariant determination to SL accuracy is independent of
both the soft breaking terms and obviously of the 
renormalization scheme. 
Ambiguities with respect to the renormalization
of $\tan \beta$ \cite{stock} would enter, however, at the single logarithmic level
at two loops.
It must also be noted that a one loop treatment is insufficient for energies above 1 TeV
in the MSSM and the SM and further investigations toward an all orders SL analysis
of the full MSSM particle production at a future linear collider is ongoing \cite{bmrv2}.

\end{document}